\documentclass{webofc}
\usepackage[varg]{txfonts}   
\usepackage{bm}
\begin{document}
\title{Electromagnetic properties of neutrinos: three new phenomena in neutrino spin oscillations}
%
%

\author{\firstname{Alexander} \lastname{Studenikin}\inst{1,2}\fnsep\thanks{\email{studenik@srd.sinp.msu.ru}}
        }

\institute{Department of Theoretical Physics, Faculty of Physics, Lomonosov Moscow State University, Moscow 119991,
Russia
\and
           Joint Institute for Nuclear Research, Dubna 141980, Moscow
Region,
Russia
          }

\abstract{%
  In studies of neutrino electromagnetic properties we discuss three very interesting aspects related to neutrino spin oscillations. First we consider neutrino mixing and oscillations in the mass and flavour bases under the influence of a constant magnetic field with nonzero transversal and longitudinal components. Then we discuss the effect of neutrino spin oscillations induced by electroweak interactions of neutrino with moving matter in case there is matter transversal  current or polarization. In the final part of the paper we discuss recently developed approach to description of neutrino spin and spin-flavour oscillations in a constant magnetic field that is based on the use of the exact neutrino stationary states in the magnetic field.
}
\maketitle

\section{Introduction}
${\bf \alpha},  {\bf \beta}=1,2$
It is well known that massive neutrinos have nontrivial electromagnetic properties, and at least the magnetic moment is not zero \cite{Fujikawa:1980yx}. Thus, neutrinos do participate also in the electromagnetic interaction (see \cite{Giunti:2014ixa} for a review). The best terrestrial laboratory upper bound on neutrino magnetic moments is obtained by the GEMMA reactor neutrino experiment \cite{Beda:2012zz}. The best astrophysical upper bound was derived from considering stars cooling \cite{Raffelt:1990pj}. The neutrino magnetic moment procession in the transversal magnetic field ${\bf B}_{\perp}$ was first considered in \cite{Cisneros:1970nq}, then spin-flavor precession in vacuum was discussed in \cite{Schechter:1981hw}, the importance of the matter effect was emphasized in \cite{Okun:1986na}. The effect of resonant amplification of neutrino spin oscillations in ${\bf B}_{\perp}$ in the presence of matter was proposed in \cite{Akhmedov:1988uk,Lim:1987tk}, the impact of the longitudinal magnetic field ${\bf B}_{||}$ was discussed in \cite{Akhmedov:1988hd}.

Here below we discuss three very interesting aspects related to the neutrino spin and spin-flavour oscillations:

1) we consider in details \cite{Fabbricatore:2016nec} neutrino mixing and oscillations in arbitrary constant magnetic field that have  ${\bf B}_{\perp}$ and ${\bf B}_{||}$ nonzero components in mass and flavour bases ,

2) we show that neutrino spin and spin-flavour oscillations can be induced not only by the neutrino interaction with a  magnetic field but also by neutrino interactions with matter in the case when there is a transversal matter current or matter polarization \cite{Studenikin:2004bu, Studenikin:2004tv} (see \cite{Studenikin:2016ykv} for historical notes and further references).

3) we develop a new (and more precise than the usual one) approach to description of neutrino spin and spin-flavor oscillations in the presence of an arbitrary magnetic field; our approach \cite{Dmitriev:2015ega} is based on the use of the stationary states in the magnetic field for classification of neutrino spin states, contrary to the customary approach when the neutrino helicity states are used for this purpose.

\section{Neutrino spin oscillations in mass and flavour bases}
\subsection{Oscillations in mass basis}

We start \cite{Fabbricatore:2016nec} with two neutrino physical states $\nu_1$ and $\nu_2$ having masses $m_1$ and $m_2$ and introduce neutrino electromagnetic interaction via magnetic moment matrix $\mu_{\alpha \beta}$, $\alpha,  \beta=1,2$:
\begin{equation}\label{a}
H_{EM} = \frac{1}{2}\mu_{\alpha \beta}\overline{\nu}_{\beta}\sigma_{\mu \nu}\nu_{\alpha}F^{\mu \nu} + h.c. \ ,
\end{equation}
where $F^{\mu \nu}$ is the electromagnetic field tensor and $\sigma_{\mu \nu}=i/2 (\gamma_{\mu}\gamma_{\nu}-\gamma_{\nu}\gamma_{\mu})$. In a uniform magnetic field the Hamiltonian (\ref{a}) becomes
\begin{equation}
\label{aa}
H_{EM} = -\mu_{\alpha \alpha '}
\bar{\nu}_{\alpha '}{ \bf{\Sigma} \bf{B}}\nu_{\alpha} + h.c. ,\ \ \ \ \Sigma_i=
\left(\begin{array}{cc}
                   \sigma_i & 0 \\
                   0 & \sigma_i
                 \end{array}\right),
\end{equation}
where
$\sigma_i$ are the Pauli matrices.

In the neutrino oscillation framework, one is interested in evolution of chiral neutrino components within the common neutrino beam state. Since in the ultrarelativistic limit the latter are approximated by free neutrino states with definite helicity $s=\pm 1$ the 4-component basis $(\nu_{1, s=1}, \nu_{1, s=-1}, \nu_{2, s=1}, \nu_{2, s=-1})$ is adopted to describe neutrino beam. With the standard column vector notation, $\nu_m \equiv (\nu_{1, s=1}, \nu_{1, s=-1}, \nu_{2, s=1}, \nu_{2, s=-1})^{\tau}$ the neutrino evolution equation relevant to electromagnetic interaction has the Schr$\ddot{o}$dinger-like form
\begin{equation}\label{schred_eq}
	i\frac{d}{dt} \nu_m (t)=H_{eff}\nu_m (t).
\end{equation}
The effective Hamiltonian consists of the vacuum and interaction parts
\begin{equation}\label{H_eff}
  H_{eff}=H_{vac}+H_{B}
\end{equation}
where the interaction part is composed with matrix elements of the field interaction Hamiltonian taken over the helicity neutrino states: $H_B=\langle	 {\nu_{\alpha,s}}|H_{EM}|{\nu_{\alpha ', s'}}\rangle$.

Let us calculate the effective interaction Hamiltonian under assumption that neutrino moves along the $z$-axis. From the magnetic field interaction Hamiltonian (\ref{a}) we have:
\begin{equation}\label{HB}
H^B_{\alpha,s; \alpha ', s'}=\langle	{\nu_{\alpha,s}}|H_{EM}|{\nu_{\alpha ', s'}}\rangle=-\frac{\mu_{\alpha, \alpha '}}{2}\int{d^3x} \nu_{\alpha}^{\dagger}\gamma_0\begin{pmatrix}{{\bf\sigma} {\bf B}} & 0 \\ 0 & \bf{\sigma} \bf{B}\end{pmatrix}\nu_{\alpha '}.
\end{equation}
In the spinor representation the free neutrino states are given by
\begin{equation}	 \label{wave func}
\nu_{\alpha,s}=C_{\alpha}\sqrt{\frac{E_{\alpha}+
m_{\alpha}}{2E_{\alpha}}}\begin{pmatrix}u_{s} \\ \frac{\bf{\sigma p_{\alpha}}}{E_{\alpha}+m_{\alpha}}u_{s}\end{pmatrix}e^{i{\bf{p}}_{\alpha} x},
\end{equation}
where ${\bf{p}}_{\alpha}$ is the neutrino $\nu_{\alpha}$ momentum. The two-component spinors $u_s$ define neutrino helicity states, and are given by
\begin{equation}\label{u_s_plus}
	u_{s=1}=\begin{pmatrix}1 \\ 0\end{pmatrix}, \ \ \ \
	u_{s=-1}=\begin{pmatrix}0 \\ 1\end{pmatrix}.
\end{equation}
Recall that in the ultrarelativistic limit these are correspondent to the right-handed $\nu_{R}$ and left-handed $\nu_{L}$ chiral neutrinos, respectively.

Substituting (\ref{wave func}) into the effective Hamiltonian formula (\ref{HB}) we get
\begin{multline}\label{HB2} H^B_{\alpha,s; \alpha ', s'}=-\frac{1}{2}\mu_{{\alpha \alpha '}}
C_{\alpha}C_{\alpha '}\int{d^3x\bm{B}\begin{pmatrix}u_{s}^{\dagger} & \frac{{\bm{\sigma p}_{\alpha}}}{E_{\alpha}+m_{\alpha}}u_{s}^{\dagger}\end{pmatrix} \begin{pmatrix}\bm{\sigma} & 0 \\ 0 & -\bm{\sigma}\end{pmatrix}\begin{pmatrix}u_{s'} \\ \frac{{\bm{\sigma p}_{\alpha '}}}{E_{\alpha '}+m_{\alpha '}}u_{s'}\end{pmatrix}}\\
\\ \times \frac{\sqrt{\left(E_{\alpha}+m_{\alpha}\right)\left(E_{\alpha '}+m_{\alpha '}\right)}}{2\sqrt{E_{\alpha}E_{\alpha '}}} \exp\left(i\Delta px\right) .\end{multline}
Decomposing the magnetic field vector into longitudinal and transversal with respect to neutrino motion components ${\bf B}={\bf B}_{||}+{\bf B}_{\perp}$ it is possible to show that
\begin{multline}
\bf{B}\begin{pmatrix}u_{s}^{\dagger} & \frac{\bm{\sigma p_{\alpha}}}{E_{\alpha}+m_{\alpha}}u_{s}^{\dagger}\end{pmatrix} \begin{pmatrix}\bm{\sigma} & 0 \\ 0 & -\bm{\sigma}\end{pmatrix}\begin{pmatrix}u_{s'} \\ \frac{\bm{\sigma p_{\alpha '}}}{E_{\alpha '}+m_{\alpha '}}u_{s'}\end{pmatrix}=\\
\\u_{s}^{\dagger} \left( \bm{\sigma B_{||}}\left(1-\frac{\bm{p_{\alpha} p_{\alpha '}}}{\left(E_{\alpha}+m_{\alpha}\right)\left(E_{\alpha '}+m_{\alpha '}\right)}\right)+\bm{\sigma B_{\perp}}\left(1+\frac{\bm{p_{\alpha} p_{\alpha '}}}{\left(E_{\alpha}+m_{\alpha}\right)\left(E_{\alpha '}+m_{\alpha '}\right)} \right) \right)  u_{s'}.
\end{multline}

Let us apply the ultrarelativistic condition $\frac{m_{\alpha}}{E_{\alpha}} \ll 1$ to the part of the integrand in (\ref{HB2}):
\begin{equation}\label{b}
\left(1 \mp \frac{\bm{p_{\alpha} p_{\alpha '}}}{\left(E_{\alpha}+m_{\alpha}\right)\left(E_{\alpha '}+m_{\alpha '}\right)}\right)\frac{\sqrt{\left(E_{\alpha}+m_{\alpha}\right)\left(E_{\alpha '}+m_{\alpha '}\right)}}{2\sqrt{E_{\alpha}E_{\alpha '}}} \approx
\left \{
  \begin{tabular}{ccc}
  $\gamma_{\alpha \alpha '}^{-1}$ \\
  $1$
    \end{tabular}
    \right.
\end{equation}
where $\gamma_{\alpha \alpha '}= \frac{1}{2}\left(\frac{m_{\alpha}}{E_{\alpha}}+\frac{m_{\alpha '}}{E_{\alpha '}}\right)$ is the transition gamma-factor.

Introducing an angle $\beta$ between $\bf{B}$ and $\bf{p}_{\alpha}$ vectors and assuming that $\bf{B_{\perp}}$ is aligned along the $x$-axis we further obtain:
\begin{equation}
u^{\dagger}_{s=1}\bm{\sigma}{\bf B}u_{s=1}=B\cos \beta , \ \ \ u^{\dagger}_{s=1}\bm{\sigma}{\bf B}u_{s=-1}=B\sin \beta ,\end{equation}
\begin{equation}u^{\dagger}_{s=-1}\bm{\sigma}{\bf B}u_{s=1}=B\sin \beta , \ \ \ u^{\dagger}_{s=-1}\bm{\sigma}{\bf B}u_{s=-1}=-B\cos \beta .\end{equation}
As it was expected, in neutrino transitions without change of helicity only the $B_{\parallel}=B\cos \beta$ component of the magnetic field contribute to the effective potential, whereas in transitions with change of the neutrino helicity the transversal component $B_{\perp}=B\sin \beta$ matters.

Performing remaining some simple algebra one can readily write out the $H_{B}$ matrix. For the effective Hamiltonian with the diagonal vacuum part $H_{vac}$ we get
\begin{equation}\label{gen_evol_eq}
	i\frac{d}{dt} \begin{pmatrix}\nu_{1, s=1} \\ \nu_{1, s=-1} \\  \nu_{2, s=1 } \\ \nu_{2, s=-1 }\end{pmatrix}=
	\begin{pmatrix}
	E_{1}+\mu_{1 1}\frac{B_{||}}{\gamma_{1 1}} & \mu_{1 1}B_{\perp} & \mu_{1 2 }
\frac{B_{||}}{\gamma_{1 2}} & \mu_{1 2 }B_{\perp} \\
	\mu_{1 1}B_{\perp} & E_{1}-\mu_{1 1}\frac{B_{||}}{\gamma_{1 1}} & \mu_{1 2}B_{\perp} & -
\mu_{1 2}\frac{B_{||}}{\gamma_{1 2 }} \\
	\mu_{1 2 }\frac{B_{||}}{\gamma_{1 2 }} & \mu_{1 2 }B_{\perp} & E_{2}+\mu_{2  2 }\frac{B_{||}}
{\gamma_{2  2 }} & \mu_{2  2 }B_{\perp} \\
	\mu_{1 2}B_{\perp} & -\mu_{1 2}\frac{B_{||}}{\gamma_{1 2 }} & \mu_{2  2}B_{\perp} & E_{2}-
\mu_{2  2 }\frac{B_{||}}{\gamma_{2  2 }}\\
	\end{pmatrix}
	\begin{pmatrix}\nu_{1, s=1} \\ \nu_{1, s=-1} \\  \nu_{2, s=1 } \\ \nu_{2, s=-1}\end{pmatrix}.
\end{equation}
This equation governs all possible oscillations of the four neutrino mass states determined by the masses $m_1$ and $m_2$ and helicities $s=1$ and $s=-1$ in the presence of a magnetic field.
Thus, it follows that:
1) the change of helicity is due to the magnetic (or transition) moment interaction with ${\bf B}_{\perp}$,
2) the longitudinal  field ${\bf B}_{||}$,  coupled to the magnetic moment, shifts the neutrino energy,
3) an additional mixing between neutrino states with different masses is induced by the magnetic moment interaction with ${\bf B}_{||}$.

\subsection{Oscillations in flavour basis}
Once having physics in the mass basis in hands, our next step is to bring it to observational terms \cite{Fabbricatore:2016nec}. This means that we must elaborate a
generalization of the mixing matrix for transitions between neutrino vector written in two four-component bases $\nu_m$ and $\nu_{f}=(\nu_{e}^{R}, \nu_{e}^{L}, \nu_{\mu}^{R}, \nu_{\mu}^{L})^{\tau}$ so that
\begin{equation}\label{U_def}
  \nu_{f}=U\nu_m.
\end{equation}
This procedure appears to be not quite direct since we should hold the condition that the polarization of the fields must preserve under transformation of the bases elements. That is why we put (still keeping in mind that chiral components are almost helicity ones):
\begin{equation}\label{transformations}
\nu_{e}^{R,L} =\nu_{1,s=\pm 1}\cos\theta+\nu_{2,s=\pm 1}\sin\theta, \ \ \
\nu_{\mu}^{R,L}=-\nu_{1,s=\pm 1}\sin\theta+\nu_{2,s=\pm 1}\cos\theta.
\end{equation}
Then, using Eqs. (\ref{U_def}) and (\ref{transformations}), it is easy to obtain that
\begin{equation}\label{U}
  U=
\begin{pmatrix}
    \cos\theta & 0 & \sin\theta & 0 \\
	0 & \cos\theta & 0 & \sin\theta \\
	-\sin\theta & 0 & \cos\theta & 0 \\
    0 & -\sin\theta & 0 & \cos\theta \\
\end{pmatrix}.
\end{equation}

Given the transition matrix (\ref{U}), derivation of the evolution equation in the flavor basis is straightforward:
\begin{equation}\label{schred_eq_fl}
  i\dfrac{d}{dt}\nu_{f}=UHU^{\dag}\nu_{f},
\end{equation}
so that the effective interaction Hamiltonian ${\tilde H}_B^{f} \equiv U H_B U^{\dag}$ has the following structure,
\begin{equation}\label{H_B_f}
{\tilde H}_B^{f}=
\begin{pmatrix}
(\frac{\mu}{\gamma})_{ee}{B_{||}}  & \mu_{ee} B_{\perp} & (\frac{\mu}{\gamma})_{e\mu}{B_{||}}  & \mu_{e\mu} B_{\perp} \\
\mu_{ee} B_{\perp} & -(\frac{\mu}{\gamma})_{ee}{B_{||}}  & \mu_{e\mu} B_{\perp} & - (\frac{\mu}{\gamma})_{e\mu}{B_{||}}  \\
(\frac{\mu}{\gamma})_{e\mu}{B_{||}}  & \mu_{e\mu} B_{\perp} & (\frac{\mu}{\gamma})_{\mu\mu}{B_{||}}  & \mu_{\mu\mu} B_{\perp} \\
\mu_{e\mu} B_{\perp} & -(\frac{\mu}{\gamma})_{e\mu}{B_{||}}  & \mu_{\mu\mu} B_{\perp} & - (\frac{\mu}{\gamma})_{\mu\mu}{B_{||}}
\end{pmatrix}.
\end{equation}
Here we introduce the following formal notations intended to manifest an analogy with the standard spin and spin-flavor oscillation formalism,
\begin{eqnarray}\label{mu_Gammas_fl}
  \nonumber \Big(\frac{\mu}{\gamma}\Big)_{ee} &=& \frac{\mu_{11}}{\gamma_{11}}\cos^2\theta+\frac{\mu_{22}}{\gamma_{22}}\sin^2\theta+\frac{\mu_{12}}{\gamma_{12}}\sin2\theta \\  \Big(\frac{\mu}{\gamma}\Big)_{e\mu} &=& \frac{\mu_{12}}{\gamma_{12}}\cos2\theta+\frac{1}{2}\Big(\frac{\mu_{22}}{\gamma_{22}}-\frac{\mu_{11}}{\gamma_{11}}
  \Big)\sin2\theta \\
  \nonumber \Big(\frac{\mu}{\gamma}\Big)_{\mu\mu} &=& \frac{\mu_{11}}{\gamma_{11}}\cos^2\theta+\frac{\mu_{22}}{\gamma_{22}}\sin^2\theta-\frac{\mu_{12}}
  {\gamma_{12}}\sin2\theta
\end{eqnarray}
\begin{eqnarray}\label{mu_fl}
  \mu_{ee} &=& \mu_{11}\cos^2\theta+\mu_{22}\sin^2\theta+\mu_{12}\sin2\theta \nonumber \\
  \mu_{e\mu} &=& \mu_{12}\cos2\theta+\frac{1}{2}(\mu_{22}-\mu_{11})\sin2\theta \\
  \mu_{\mu\mu} &=& \mu_{11}\cos^2\theta+\mu_{22}\sin^2\theta-\mu_{12}\sin2\theta \nonumber
\end{eqnarray}
It should be noted that equations (\ref{mu_fl}) follows from the general expression that settles relations between two neutrino bases \cite{Giunti:2014ixa}.

It is interesting to consider a particular case when only the longitudinal magnetic field ${{\bf B}_{||}}$ is present in an environment (${\bf B}_{\perp}=0$). Then the Hamiltonian (\ref{H_B_f}) is reduced to the following one,
\begin{equation}\label{H_B_f_decopled}
{\tilde H}_B^{f}=
\begin{pmatrix}
(\frac{\mu}{\gamma})_{ee}{B_{||}}  & 0 & (\frac{\mu}{\gamma})_{e\mu}{B_{||}}  & 0 \\
0 & -(\frac{\mu}{\gamma})_{ee}{B_{||}}  & 0 & - (\frac{\mu}{\gamma})_{e\mu}{B_{||}}  \\
(\frac{\mu}{\gamma})_{e\mu}{B_{||}}  & 0 & (\frac{\mu}{\gamma})_{\mu\mu}{B_{||}}  & 0 \\
0 & -(\frac{\mu}{\gamma})_{e\mu}{B_{||}}  & 0 & - (\frac{\mu}{\gamma})_{\mu\mu}{B_{||}}
\end{pmatrix}.
\end{equation}

Obviously, the neutrino states with different flavor and same chirality decouple and form subsystems independently mixed by the magnetic field. For example, one would have two states $(\nu^L_{e}, \nu^L_{\mu})$ mixed in accordance with the equation
\begin{equation}\label{3_evol_eq}
	i\frac{d}{dt} \begin{pmatrix}\nu^L_{e} \\ \nu^L_{\mu} \\  \end{pmatrix}=
	\begin{pmatrix}
	-\frac{\Delta m^2}{4E}\cos 2\theta - (\frac{\mu}{\gamma})_{ee}{B_{||}}&  \frac{\Delta m^2}{4E}\sin 2\theta - (\frac{\mu}{\gamma})_{e\mu}{B_{||}}  \\
	 \frac{\Delta m^2}{4E}\sin 2\theta - (\frac{\mu}{\gamma})_{e\mu}{B_{||}} & \frac{\Delta m^2}{4E}\cos 2\theta - (\frac{\mu}{\gamma})_{\mu\mu}{B_{||}}  \\
		\end{pmatrix}
	\begin{pmatrix}\nu^L_{e} \\ \nu^L_{\mu} \\ \end{pmatrix}.
\end{equation}
From this it follows that the neutrino magnetic moment interactions with the longitudinal magnetic field can generate the neutrino flavour mixing (an additional mixing to the usual effect due to neutrino mixing angle $\theta$) without changing neutrino chirality. For the flavour neutrino oscillation probability in the adiabtic case we get
\begin{equation}
P_{\nu_{e}^{L}\rightarrow\nu_{\mu}^{L}}=\frac{\Big( \frac{\Delta m^2}{2E}\sin 2\theta - 2(\frac{\mu}{\gamma})_{e\mu}{B_{||}}\Big)^2}{{\Big( \frac{\Delta m^2}{2E}\sin 2\theta - 2(\frac{\mu}{\gamma})_{e\mu}{B_{||}}\Big)^2}+{\Big( \frac{\Delta m^2}{2E}\cos 2\theta + 2\frac{\mu_{12}}{\gamma_{12}}{B_{||}}\sin 2\theta\Big)^2}}\sin^2\Big(\frac{1}{2}{\sqrt D} x\Big),
\end{equation}
where
\begin{equation}
D={{\Big( \frac{\Delta m^2}{2E}\sin 2\theta - 2(\frac{\mu}{\gamma})_{e\mu}{B_{||}}\Big)^2}+{\Big( \frac{\Delta m^2}
{2E}\cos 2\theta + 2\frac{\mu_{12}}{\gamma_{12}}{B_{||}}\sin 2\theta\Big)^2}}.
\end{equation}
It follows that $B_{||}$ not only generates flavour neutrino mixing but also can produce the resonance amplification of the corresponding oscillations.

\section{Neutrino spin precession and oscillations due to matter transversal motion}

 Consider, as an example,  an electron neutrino spin procession in the case when neutrinos with the Standard Model interaction are propagating through moving and polarized matter composed of electrons (electron gas) in the presence of an electromagnetic field given by the electromagnetic-field tensor $F_{\mu \nu}=({\bf E}, {\bf B})$. As discussed in \cite{Studenikin:2004bu, Studenikin:2004tv}
(see also \cite{Egorov:1999ah,Lobanov:2001ar, Dvornikov:2002rs}), the the generalized Bargmann-Michel-Telegdi equation describes  the evolution of the
three-di\-men\-sio\-nal neutrino spin vector $\vec S $,
\begin{equation}\label{S}
{d{\bf S} \over dt}={2\mu \over \gamma} \Big[ {\bf S} \times ({\bf
B}_0+{\bf M}_0) \Big],
\end{equation}
where the magnetic field $\bf{B}_0$ in the neutrino rest frame is determined by the transversal
and longitudinal (with respect to the neutrino motion) magnetic and electric field components in the
laboratory frame,
\begin{equation}
{\bf  B}_0=\gamma\Big({\bf B}_{\perp} +{1 \over \gamma} {\bf
B}_{\parallel} + \sqrt{1-\gamma^{-2}} \Big[ {\bf E}_{\perp} \times
\frac{{\bm\beta}}{\beta} \Big]\Big),
\end{equation}
$\gamma = (1-\beta^2)^{-{1 \over 2}}$, $\bm{\beta}$ is the neutrino velocity.

The matter term ${\bf M}_0$ in Eq. (\ref{S}) is also composed of the transversal ${\bf  M}{_{0_{\parallel}}}$
and longitudinal  ${\bf  M}_{0_{\perp}}$ parts,
\begin{equation}
{\bf M}_0=\bf {M}{_{0_{\parallel}}}+{\bf M}_{0_{\perp}},
\label{M_0}
\end{equation}
\begin{equation}
\begin{array}{c}
\displaystyle {\bf M}_{0_{\parallel}}=\gamma{\bf \beta}{n_{0} \over
\sqrt {1- v_{e}^{2}}}\left\{ \rho^{(1)}_{e}\left(1-{{\bf v}_e
{\bm\beta} \over {1- {\gamma^{-2}}}} \right)\right\}-
\displaystyle{\rho^{(2)}_{e}\over {1- {\gamma^{-2}}}} \left\{{\bm\zeta}_{e}{\bm\beta}
\sqrt{1-v^2_e}+ {\left({\bm \zeta}_{e}{{\bf v}_e}\frac{{\bm\beta}{\bf v}_e}{{1+\sqrt{1-v^2_e}} }\right)}
\right\}, \label{M_0_parallel}
\end{array}
\end{equation}
\begin{equation}\label{M_0_perp}
\begin{array}{c}
\displaystyle {\bf M}_{0_{\perp}}=-\frac{n_{0}}{\sqrt {1-
v_{e}^{2}}}\Bigg\{ {\bf v}_{e_{\perp}}\Big(
\rho^{(1)}_{e}+\rho^{(2)}_{e}\frac
{{\bm\zeta}_{e} {\bf v}_e} {1+\sqrt{1-v^2_e}}\Big) +
\displaystyle {{\bm\zeta}_{e_{\perp}}}\rho^{(2)}_{e}\sqrt{1-v^2_e}\Bigg\}.
\end{array}
\end{equation}
Here $n_0=n_{e}\sqrt {1-v^{2}_{e}}$ is the invariant number density of
matter given in the reference frame for which the total speed of
matter is zero. The vectors ${\bf v}_e$, and ${\bm \zeta}_e \
(0\leq |{\bm \zeta}_e |^2 \leq 1)$ denote, respectively,
the speed of the reference frame in which the mean momentum of
matter (electrons) is zero, and the mean value of the polarization
vector of the background electrons in the above mentioned
reference frame. The coefficients $\rho^{(1,2)}_e$ calculated
within the extended Standard Model supplied with $SU(2)$-singlet right-handed neutrino
$\nu_{R}$ are respectively,  $\rho^{(1)}_e={\tilde{G}_F \over {2\sqrt{2}\mu }}, \ \ \rho^{(2)}_e =-{G_F \over {2\sqrt{2}\mu}}$,
where $\tilde{G}_{F}={G}_{F}(1+4\sin^2 \theta _W).$
For neutrino evolution between two neutrino states $\nu_{e}^{L}\Leftrightarrow\nu_{e}^{R}$ in presence of the magnetic field and moving matter we get the following equation
\begin{equation}\label{2_evol_eq}
	i\frac{d}{dt} \begin{pmatrix}\nu_{e}^{L} \\ \nu_{e}^{R} \\  \end{pmatrix}={\mu}
	\begin{pmatrix}
	{1 \over \gamma}\big|{\bf
M}_{0\parallel}+{{\bf B}}_{0\parallel}\big| & \big|{{\bf B}}_{\perp} + {1\over
\gamma}{\bf M}_{0\perp} \big|  \\
	 \big|{{\bf B}}_{\perp} + {1\over
\gamma}{\bf M}_{0\perp} \big| & -{1 \over \gamma}\mid{\bf
M}_{0\parallel}+{{\bf B}}_{0\parallel}\big|  \\
		\end{pmatrix}
	\begin{pmatrix}\nu_{e}^{L} \\ \nu_{e}^{R} \\ \end{pmatrix}.
\end{equation}
Thus, the probability of the neutrino spin oscillations in the adiabatic
approximation is given by (see \cite{Studenikin:2004bu, Studenikin:2004tv})
\begin{equation}\label{ver2}
P_{\nu_L \rightarrow \nu_R} (x)=\sin^{2} 2\theta_\textmd{eff}
\sin^{2}{\pi x \over L_\textmd{eff}},\ \sin^{2} 2\theta_\textmd{eff}={E^2_\textmd{eff} \over
{E^{2}_\textmd{eff}+\Delta^{2}_\textmd{eff}}}, \ \ \
L_\textmd{eff}={2\pi \over
\sqrt{E^{2}_\textmd{eff}+\Delta^{2}_\textmd{eff}}},
\end{equation}
where
$E_\textmd{eff}=\mu \big|{{\bf B}}_{\perp} + {1\over
\gamma}{\bf M}_{0\perp} \big|, \ \
\Delta_ \textmd{eff}={\mu \over \gamma}\big|{\bf
M}_{0\parallel}+{{\bf B}}_{0\parallel} \big|$.
It follows that even without presence of an electromagnetic field,
${{\bf B}}_{\perp}={{\bf B}}_{0\parallel}=0$,
neutrino spin  oscillations can be induced in the presence of matter
when the transverse matter term ${\bf M}_{0\perp}$ is not zero.
This possibility is realized in the case when the transverse component of the background matter velocity or its transverse polarization is not zero. It is obvious that for neutrinos
with nonzero transition magnetic moments a similar effect of spin-flavour
oscillations exists under the same background conditions. A possibility of neutrino
spin procession and oscillations induced by the transversal matter current or polarization
was first discussed in \cite{Studenikin:2004bu, Studenikin:2004tv}.
The existence of this effect has been recently confirmed  in
\cite{Kartavtsev:2015eva,Volpe:2015rla,Cirigliano:2014aoa} where neutrinos
propagation in anisotropic media is studied.

\section{Neutrino spin oscillations and stationary spin states in magnetic field}
\label{Sec_4}

We develop a new  approach \cite{Dmitriev:2015ega}, that is more precise than the usual one, to description of neutrino spin and spin-flavor oscillations in the presence of an arbitrary magnetic field. The derived probability of neutrino oscillations does not coincide with the usual one and the difference might have important phenomenological consequences.  Within this customary approach the helicity operator is used for classification of a neutrino spin states in a magnetic field. However, the helicity operator does not commute with the neutrino evolution Hamiltonian in a magnetic field. This case resembles situation of the flavour neutrino oscillations in the nonadiabatic case when the neutrino mass states are not stationary.  The proposed alternative approach to neutrino spin oscillations is based on the exact solutions of the corresponding Dirac equation for a massive neutrino wave function in the presence of a magnetic field that stipulates the description of the neutrino spin states with the corresponding spin operator that commutes with the neutrino dynamic Hamiltonian in the magnetic field.

Here we again consider a simple model with two generations of flavour neutrinos
$\nu_e$ and $\nu_\mu$ that are the orthogonal superpositions of mass states $\nu_1$ and $\nu_2$
$\nu_f=\sum_iU_{fi}\nu_i,$
where $U_{fi}$ are elements the mixing matrix given by (\ref{transformations}) and $f=e,\mu$, $i=1,2$.
We start with consideration of a massive $\nu_i$ with the magnetic moment $\mu_i$ that propagates  along $\bf{n}_{z}$ direction in presence of constant homogeneous arbitrary orientated magnetic field ${\bf B}=(B_\bot,0,B_\|)$. The neutrino wave function in the momentum representation is given by a plane wave solution of the modified Dirac equation
\begin{equation}\label{eq1}
  (\gamma p-m_i-{\mu_i}{\bm{\Sigma}\bf{B}})\nu_i(p)=0.
\end{equation}
The neutrino energy spectrum can be determined from the condition
\begin{equation}\det(\gamma p -m_i-{\mu_i}{\bm{\Sigma}\bf{B}})=0,\end{equation}
which guarantees the existence of a nontrivial solution of the modified Dirac equation (\ref{eq1}).
For the neutrino energy spectrum we obtain
\begin{equation}\label{spec}
E_i^\pm=\sqrt{m_i^2+p^2+{\mu_i}^2{\bf{B}}^2\pm2{\mu_i}\sqrt{m_i^2{\bf{B}}^2+p^2B_\bot^2}},
\end{equation}
where ``$\pm$'' denotes two different eigenvalues of the Hamiltonian
$\label{Ham_i}
  H_i=\gamma_0(m_i+{\bm{\gamma}\bf{p}}+{\mu_i}{\bm{\sigma}\bf{B}})$,
which describes dynamics of the neutrino system under consideration.

We define different neutrino spin states in the mass basis as eigenstates
of the spin operator
\begin{equation}\label{spin}
S_i=\frac{1}{|{\bf{B}}|}({\bf{\Sigma}\bf{B}}-\frac{i}{m_i}
\gamma_0\gamma_5\left[{\bf{\Sigma}}\times{\bf{p}}\right]{\bf{B}}),
\end{equation}
which commutes with the Hamiltonian $H_i$. Hence, we specify the neutrino spin states
as the stationary states for the Hamiltonian, contrary to the case when the helicity operator is used.

Consider the mass state $\nu_{i}$ as a superposition of neutrinos $\nu_i^+$ and $\nu_i^-$ in a
definite spin state, $\nu_i=c_i^+\nu_i^++c_i^-\nu_i^-$.
The complex coefficients denote two different eigenstates of the spin operator $S_i$ and  $|c_i^+|^2+|c_i^-|^2=1$. Thus, the neutrino mass states evolve
following to
\begin{equation}\label{nui}
\nu_i(x)=\left[c_i^+e^{-iE_i^+t}\xi_i^++c_i^-e^{-iE_i^-t}\xi_i^-\right]e^{i\bm {p}\bm {x}},
\end{equation}
where the neutrino initial state at $t=0$ is given by
$\nu_i^\pm (t=0)= \xi_i^\pm e^{i\bm {p}\bm {x}}$.
In the following calculations the term $e^{i\bm {p}\bm {x}}$ is neglected because it is
irrelevant for the neutrino oscillation probability.

Next we assume that the initial neutrino state $\nu (t=0)$ is a pure electron state
which is defined as the superposition of the mass states,
\begin{equation}\label{nu_0}
\nu (t=0) =\left[c_1^+\xi_1^++c_1^-\xi_1^-\right]\cos\theta
+\left[c_2^+\xi_2^++c_2^-\xi_2^-\right]\sin\theta.
\end{equation}
Using (\ref{nui}) we see that this state depends on time as
\begin{equation}\label{nu_t}\nu(t)=\left[c_1^+e^{-iE_1^+t}\xi_1^++c_1^-e^{-iE_1^-t}\xi_1^-\right]\cos\theta
+\left[c_2^+e^{-iE_2^+t}\xi_2^++c_2^-e^{-iE_2^-t}\xi_2^-\right]\sin\theta.\end{equation}
Therefore, the probability to observe the muon neutrino state $\nu_\mu$ at time $t$ is given by
$P_{\nu_e\rightarrow\nu_\mu}(t)=\left| \left\langle \nu_\mu|\nu(t)\right\rangle \right|^2$,
where
$\nu_\mu=-\left[c_1^+\xi_1^++c_1^-\xi_1^-\right]\sin\theta+\left[
c_2^+\xi_2^++c_2^-\xi_2^-\right]\cos\theta.$
It is clear that
$\xi_i^{s\dag}\xi_i^{s^\prime}=\delta^
{ss^\prime} \text{and}
\ \ \
\xi_2^{s\dag}\xi_1^{s^\prime}=\xi_1^{s\dag}\xi_2^{s^\prime}=0$,
where $s,s^\prime=\pm$. Using (\ref{nu_0}) and (\ref{nu_t}) we get for the oscillation probability
\begin{multline}\label{P}
  P_{\nu_e\rightarrow\nu_\mu}(t)=\left\{-|c_2^+|^2|c_2^-|^2\sin^2\frac{E_2^+-E_2^-}{2}t+
  |c_2^+|^2|c_1^+|^2\sin^2\frac{E_2^+-E_1^+}{2}t+\right.\\+|c_2^+|^2|c_1^-|^2\sin^2\frac{E_2^+-E_1^-}{2}t
  +|c_2^-|^2|c_1^+|^2\sin^2\frac{E_2^--E_1^+}{2}t+\\\left.+|c_2^-|^2|c_1^-|^2\sin^2\frac{E_2^--E_1^-}{2}t
  -|c_1^+|^2|c_1^-|^2\sin^2\frac{E_1^+-E_1^-}{2}t\right\}\sin^22\theta.
\end{multline}

It is usually assumed that the initial state of relativistic neutrino is a negative-helicity state,
which means that
\begin{equation}\frac{\bf{\sigma}\bf{p}}{|\bf{p}|}\xi_i=-\xi_i.
            \end{equation}
Next we consider the left-handed spinors because only the left-handed
fermions participate in the production and detection processes and we suppose that each of
the mass states of the initial electron neutrino are left-handed. In our case the helicity operator
is equal to ${\bm{\sigma}\bf{p}}/{|{\bf{p}}|}=\sigma_3$, therefore the initial neutrino
state is given by $\psi_L=\left(0,1,0,0\right)^{T}$.
Let us write the initial neutrino state  $\psi_L$ as a superposition of the eigenvectors of
the spin operator $S_i$. From (\ref{spin}) we get
\begin{equation}S_i=\left(\begin{array}{cccc}
              \cos\phi & \sin\phi & 0 & -\frac{p}{m_i}\sin\phi \\
              \sin\phi & -\cos\phi & \frac{p}{m_i}\sin\phi & 0 \\
              0 & \frac{p}{m_i}\sin\phi & \cos\phi & \sin\phi \\
              -\frac{p}{m_i}\sin\phi & 0 & \sin\phi & -\cos\phi
            \end{array}\right),
\end{equation}
where $\phi$ is the angle between $\bf{B}$ and $\bf{p}$. It is obvious that
$S_i^2=\left(1+\frac{p^2}{m_i^2}\sin^2\phi\right)\hat{I}_{4\times4}$.
In order to define the spin projector operators we introduce the normalized spin operator following to
\begin{equation}\label{s_tilde}
\tilde{S}_i=\sqrt{\frac{1}{1+\frac{p^2}{m_i^2}\sin^2\phi}}S_i\equiv N_iS_i, \ \ \ \tilde{S}_i^2=1, \ \ \
N_i=\sqrt{\frac{1}{1+\frac{p^2}{m_i^2}\sin^2\phi}}.
\end{equation}
The spin projector operators are
$P_i^\pm=\frac{1\pm \tilde{S}_i}{2}$,
and we use them to split the initial neutrino state $\psi_L$
in two neutrino states with definite spin quantum numbers
\begin{equation}\psi_i^+=P_i^+\psi_L=\frac{1}{2}\left(\begin{array}{c}
                               N_i\sin\phi \\
                               1-N_i\cos\phi \\
                               \frac{p}{m_i}N_i\sin\phi \\
                               0
                             \end{array}\right), \ \ \
\psi_i^-=P_i^-\psi_L=\frac{1}{2}\left(\begin{array}{c}
                               -N_i\sin\phi \\
                               1+N_i\cos\phi \\
                               -\frac{p}{m_i}N_i\sin\phi \\
                               0
                             \end{array}\right).\end{equation}
Note that $\psi_i=\psi_i^++\psi_i^-=c_i^+\eta_i^++c_i^-\eta_i^-,$ where $\eta_i^\pm$ is a
basis in the spin operator  $S_i$ eigenspace. From the condition
$\psi_i^{\dag\pm}\psi_L^\pm=|c_i^\pm|^2$
 we get that
$|c_i^+|^2=\frac{1-N_i\cos\phi}{2},\ \ \ |c_i^-|^2=\frac{1+N_i\cos\phi}{2}$.

Now we can insert the obtained expressions for $|c_i^\pm|^2$ in \eqref{P}.
In the forthcoming evaluation of the probability $P_{\nu_e\rightarrow\nu_\mu}(t)$ we consider the case when the magnetic field $\bf{B}$ is nearly a transversal one and $B_\bot\gg B_\|$, therefore
$\sin\phi\approx1, \ \ \ \cos\phi\approx 0$.
Then we get (it is also supposed that $\frac{p^2}{m_i^2}\gg1$)
\begin{equation}N_i\approx1-\frac{\frac{p^2}{2m_i^2}\sin^2\phi}{1+\frac{p^2}{m_i^2}\sin^2\phi}
\approx\frac{m_i^2}{p\sin^2\phi}.\end{equation}
Thus, for typical combinations of the coefficients $|c_i^\pm|^2$ of eq.(\ref{P}) in the linear approximation over $\cos \phi \ll 1$ we get

  \begin{equation}|c_i^+|^2|c_i^-|^2=\frac{1-N_i^2\cos^2\phi}{4}\approx\frac{1}{4},\end{equation}
  \begin{equation}|c_2^s|^2|c_1^s|^2\approx\frac{1}{4}\left(1-s(N_1+N_2)\cos\phi\right)
       \approx
       \frac{1}{4}\left(1-s\frac{m_1^2+m_2^2}{p^2}\cos\phi\right), \end{equation}
\begin{equation}|c_2^+|^2|c_1^-|^2\approx\frac{1}{4}\left(1-(N_2-N_1)\cos\phi\right)\approx
  \frac{1}{4}\left(1-\frac{m_2^2-m_1^2}{p^2}\cos\phi\right),\end{equation}
\begin{equation}|c_2^-|^2|c_1^+|^2\approx\frac{1}{4}\left(1-(N_1-N_2)\cos\phi\right)\approx
  \frac{1}{4}\left(1-\frac{m_1^2-m_2^2}{p^2}\cos\phi\right).\end{equation}
Using these expressions finally from \eqref{P} we get
\begin{multline}\label{fin}
  P_{\nu_{eL}\rightarrow\nu_\mu}(t)=\frac{1}{4}\sin^22\theta\left\{-\sin^2\frac{E_2^+-E_2^-}{2}t+
  \sin^2\frac{E_2^+-E_1^+}{2}t+\sin^2\frac{E_2^+-E_1^-}{2}t+
  \right.\\\left.+\sin^2\frac{E_2^--E_1^+}{2}t+\sin^2\frac{E_2^--E_1^-}{2}t
  -\sin^2\frac{E_1^+-E_1^-}{2}t\right\}+\\+\frac{m_1^2+m_2^2}{p^2}\sin^22\theta\cos\phi\left\{
  \sin^2\frac{E_2^--E_1^-}{2}t-\sin^2\frac{E_2^+-E_1^+}{2}t\right\}+\\+\frac{m_2^2-m_1^2}{p^2}\sin^22\theta\cos\phi\left\{\sin^2\frac{E_2^--E_1^+}{2}t-\sin^2\frac{E_2^+-E_1^-}{2}t\right\}.
\end{multline}
Note that two last terms here are suppressed by the presence of $\cos \phi \ll 1$.
If we also account for a rather general condition $2\mu B_\bot \ll p$  then for the neutrino energies we get
\begin{equation}E_i^\pm\approx\sqrt{m_i^2+p^2\pm2\mu_{i} pB_\bot\left(1+\frac{m_i^2}{2p^2}\right)}\approx
p\sqrt{1+\frac{m_i^2}{p^2}\pm\frac{2\mu_{i} B_\bot}{p}}\approx p+\frac{m_i^2}{2p}\pm\mu_{i} B_\bot.\end{equation}
Finally,  for the neutrino oscillations probability in the flavour basis we get (here $\Delta m^2\equiv m_2^2-m_1^2$)
\begin{multline}\label{appP}
  P_{\nu_{eL}\rightarrow\nu_\mu}(t)\approx\sin^22\theta\sin^2\frac{\Delta m^2}{4p}t+\frac{1}{2}\left(\sin^2\frac{\mu_2-\mu_1}{2}
  B_\bot t+\sin^2\frac{\mu_2+\mu_1}{2}B_\bot t\right)
  \sin^22\theta \cos\frac{\Delta m^2}{2p}t \\-\frac{1}{4}\sin^22\theta\left(\sin^2\mu_1B_\bot t+
  \sin^2\mu_2B_\bot t\right).\end{multline}
It should be emphasized that, as it follows from the above derivations, the obtained probability $P_{\nu_{eL}\rightarrow\nu_\mu}(t)$ accounts for the transitions from the initial
left-handed electron neutrino to the final muon neutrino that can be in both left- and right-handed
states.  Note that in the mass basis the transition $\nu_{1 L}\rightarrow\nu_{2 R}$ is not possible when $\mu_{12}=0$ (see (\ref{2_evol_eq})).

Consider the difference $\Delta P(t)=P_{\nu_{eL}\rightarrow\nu_\mu}-P_{\nu_{eL}\rightarrow\nu_{\mu L}}$ of the probability $P_{\nu_{eL}\rightarrow\nu_\mu}(t)$ given by (\ref{appP})
and the usual result $P_{\nu_{eL}\rightarrow\nu_{\mu L}}(t)=\sin^22\theta\sin^2\frac{\Delta m^2}{4E_\nu}t$
for $\nu_{eL}\rightarrow\nu_{\mu L}$. Obviously, this difference is just a probability for the transition $P_{\nu_{eL}\rightarrow\nu_{\mu R}}$, and we get (here  $\Delta\mu_\pm=\mu_2\pm\mu_1$)
\begin{multline}\label{appP}
  P_{\nu_{eL}\rightarrow\nu_{\mu R}}=\frac{1}{2}\left(\sin^2\frac{\Delta\mu_-}{2}
  B_\bot t+\sin^2\frac{\Delta\mu_+}{2}B_\bot t\right)
  \sin^22\theta \cos\frac{\Delta m^2}{2p}t \\-\frac{1}{4}\sin^22\theta\left(\sin^2\mu_1B_\bot t+
  \sin^2\mu_2B_\bot t\right).
\end{multline}

  Note that in the usual approach (see, for instance, in \cite{Giunti:2014ixa})
transitions $\nu_{eL}\rightarrow\nu_{\mu R}$ in $B_{\perp}$ are not possible if the
transition magnetic moments are zero. However, within the developed approach it is shown that the previous statement is not correct.

The author is thankful to the organizers of the  19th International Seminar on High Energy Physics
QUARKS-2016 for the invitation to attend this very interesting event. This work was supported by the Russian Basic Research Foundation grants No. 14-22-03043, 15-52-53112, 16-02-01023 and 17-52-53133.

\end{document}